# Interpretation of Elasticity of Liquid Marbles

Gene Whyman[a], Edward Bormashenko[*a,b]

[a]*Ariel University, Physics Faculty, P.O.B. 3, 40700, Ariel, Israel*

[b]*Ariel University, Chemical Engineering and Biotechnology Department , P.O.B. 3, 40700, Ariel, Israel*

[*]Corresponding author:

Edward Bormashenko

Ariel University, Physics Faculty,

P.O.B. 3

Ariel 40700

Phone: +972-3-906-6134

Fax: +972-3-906-6621

E-mail: edward@ariel.ac.il

**Abstract**

Liquid marbles are non-stick droplets covered with micro-scaled particles. Liquid marbles demonstrate quasi-elastic properties when pressed. The interpretation of the phenomenon of elasticity of liquid marbles is proposed. The model considering the growth in the marble surface in the course of deformation under the conservation of marble's volume explains semi-quantitatively the elastic properties of marbles in satisfactory agreement with the reported experimental data. The estimation of the effective Young modulus of marbles and its dependence on the marble volume are reported.

1. Introduction

Liquid marbles are continuing to draw the attention of investigators [1-4]. Liquid marble, shown in Fig. 1, is the non-stick droplet encapsulated with micro- or nano-scaled solid particles [5-7]. Since liquid marbles were introduced in the pioneering works of Quèrè *et al*., they have been exposed to the intensive theoretical and experimental research [8-13]. An interest in liquid marbles arises from both their very unusual physical properties and their promising applications. Liquid marbles present an alternate approach to superhydrophobicity, i.e. creating a non-stick situation for a liquid/solid pair. Usually superhydrophobicity is achieved by a surface modification of a solid substrate. In the case of liquid marbles, the approach is opposite: the surface of a liquid is coated by particles, which may be more or less hydrophobic [13]. Marbles coated by graphite and carbon black, which are not strongly hydrophobic, were also reported [14-15].

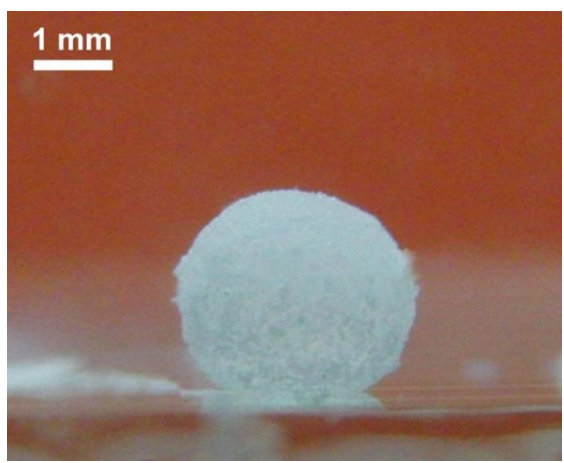

**Fig. 1.** Typical 10 µl water marble coated by polyethylene particles.

A variety of media, including organic and ionic liquids and liquid metals, could be converted into liquid marbles [16-18]. Liquid marbles were successfully exploited for microfluidics [1, 4, 18-21], water pollution detection [22], gas sensing [23], blood typing [24] and optical probing [25]. Respirable liquid marbles for the cultivation of microorganisms and Daniel cells based on liquid marbles were reported recently by Shen *et al.* [26-27]. Stimulus (pH, UV and IR) responsive liquid marbles were reported by Dupin, Fujii *et al.* [28-30]. It is noteworthy that liquid marbles retain non-stick properties on a broad diversity of solid and liquid supports [31]. Actually, liquid marbles are separated from the support by air cushions in a way similar to Leidenfrost droplets [32]. The state-of-the-art in the study of properties and applications of liquid marbles is covered in recent reviews [33-36].

Remarkably, liquid marbles demonstrate certain elastic properties and can sustain a reversible deformation of up to 60% [37]. Our paper is devoted to elucidating elastic properties of liquid marbles.

## 2. The model

We relate the stress, arising from deformation of the marble, to the growth of its surface energy due to the area increase. We accept a simple model, approximating the shape of the deformed marble by a symmetrical spherical segment, and take into account the conservation of marble volume, $V=V_0$, in the course of deformation:

$$V = \frac{4}{3}\pi R^3 - \frac{2}{3}h^2(3R - h), \qquad V_0 = \frac{4}{3}\pi R_0^3, \qquad (1)$$

where $R_0$ and $R$ are the radii of non-deformed and deformed marbles, respectively, and $h$ is the deformation (see Fig.2A). The deformations of the spherical marble from above and from below are supposed to be equal. Thus, the influence of its weight is neglected that is a reasonable simplification for small marbles, when the radius of a marble is smaller than the so-called capillary length $l_{ca} = \sqrt{\gamma_{eff}/\rho g}$, and $\gamma_{eff}$ is the effective surface tension of a marble [10]; $\rho$ is its density. The effective surface tension of liquid marbles is not single-valued due to its pronounced hysteretic nature [10]; however, for a sake of a very rough approximation a value of $\gamma_{eff} \cong 60\text{mN/m}$ may be assumed. Thus the value of the capillary length is $l_{ca} \cong 2.5\text{mm}$ for various kinds of coating powders.

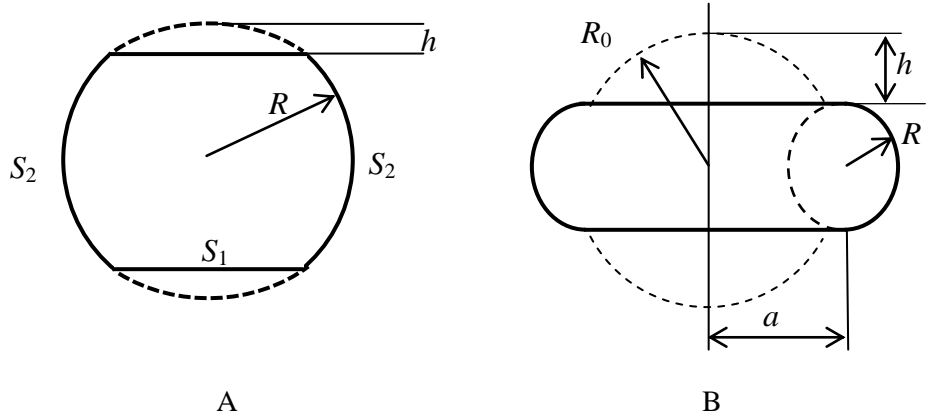

**Fig. 2. A.** The model of the deformed marble presented as a symmetric spherical segment. The upper and lower circles, which arise as a result of the equal deformations $h$, have the same area $S_1$. The side area of the spherical segment is $S_2$. In the course of deformation, the radius $R$ increases, but the segment volume is conserved. **B.** The "pancake" model of the strongly deformed marble.

The requirement of conservation of volume dictates the following expansion of $R$ in terms of powers of $h$ up to the fourth order

$$R \approx R_0 + \frac{1}{2R_0}h^2 - \frac{1}{6R_0^2}h^3 + O(h^5). \tag{2}$$

Note the absence of the first and fourth powers of $h$ from expansion (2). The overall area of the upper and lower circles is, on account of (2)

$$2S_1 = 2\pi(2Rh - h^2) \approx 2\pi\left(2R_0h - h^2 + \frac{1}{R_0}h^3 - \frac{1}{3R_0^2}h^4\right). \tag{3}$$

In the same approximation, the side area of the spherical segment is

$$S_2 = 4\pi R^2 - 4\pi Rh \approx 4\pi\left(R_0^2 - R_0h + h^2 - \frac{5}{6R_0}h^3 + \frac{5}{12R_0^2}h^4\right). \tag{4}$$

The overall increase in the surface area takes a form

$$\Delta S = 2S_1 + S_2 - 4\pi R_0^2 \approx 4\pi\left(\frac{1}{2}h^2 - \frac{1}{3R_0}h^3 + \frac{1}{4R_0^2}h^4\right). \tag{5}$$

The increase in the area under deformation induces the increase in the surface energy of the marble, $\Delta E = \gamma \Delta S$, and the corresponding elastic force is:

$$F = \frac{d\Delta E}{dh} \approx 4\pi\gamma_{\text{eff}}\left(h - \frac{1}{R_0}h^2 + \frac{1}{R_0^2}h^3\right). \tag{6}$$

where $\gamma_{\text{eff}}$ is the effective surface tension of the composite marble surface. A characteristic feature of this elastic force is the inflection, $d^2F/dh^2 = 0$, at

$$h = \frac{1}{3}R_0. \tag{7}$$

It is latently assumed, that $\gamma_{\text{eff}}$ does not change in a course of deformation. Proceeding from Eq.(6), the stress-strain dependence is determined as:

$$\sigma = \frac{4\gamma_{\text{eff}}}{R_0}(\varepsilon - \varepsilon^2 + \varepsilon^3). \tag{8}$$

where $\sigma = \frac{F}{\pi R_0^2}$ and $\varepsilon = \frac{h}{R_0}$.

Surprisingly, all the coefficients at the powers of *h* came out the same. Note the unusual even power of strain in (8) that reflects the asymmetry in pressing and stretching liquid marbles. Even if the stretching were possible experimentally, the marble would turn under stretching into a body resembling a prolate spheroid (not a spherical segment like that in Fig.1) with a different dependence of the surface area on the deformation *h*. Expression (8) also predicts a weak inverse dependence of stress on the marble volume (as the inverse of the cubic root), as well as the Young modulus in a linear approximation

$$E \approx \frac{4\gamma_{eff}}{R_0}. \tag{9}$$

Remarkably, a similar expression for the effective Young modulus of bouncing droplets was reported in Ref. 38. For small deformations the elastic Young modulus is defined by Eq. (9). The calculation gives the values of 280, 220 and 190 Pa for polyethylene marbles of the volumes of 5, 10, and 15 µL (respectively, the radii are 1.06, 1.34, and 1.53 mm).

Taking into account the nonlinear terms in Eq. (8) leads to the dependence of the Young modulus on the deformation:

$$E = \frac{4\gamma_{\text{eff}}}{R_0}(1 - \varepsilon + \varepsilon^2). \tag{9a}$$

### 3. Discussion

Recently, Sedev and coauthors obtained valuable experimental data related to marbles' deformation [37]. We compare consequences of the present model with their results. As is seen from Fig. 3, the present model reproduces the inflection point of the elastic force (6) at small deformations. The comparison shows a satisfactory correspondence of measured and calculated values, except of large deformations where, obviously, the marble shape cannot be adequately described by a spherical segment (it should also be noted that the value of the effective surface tension $\gamma_{eff}$ may change under large deformations [10]) Also expansions for the volume and surface area to a restricted order used above become less exact for relative deformations $\varepsilon = h/R_0$ approaching unity. Note, however, that the model does not include any fitting parameter.

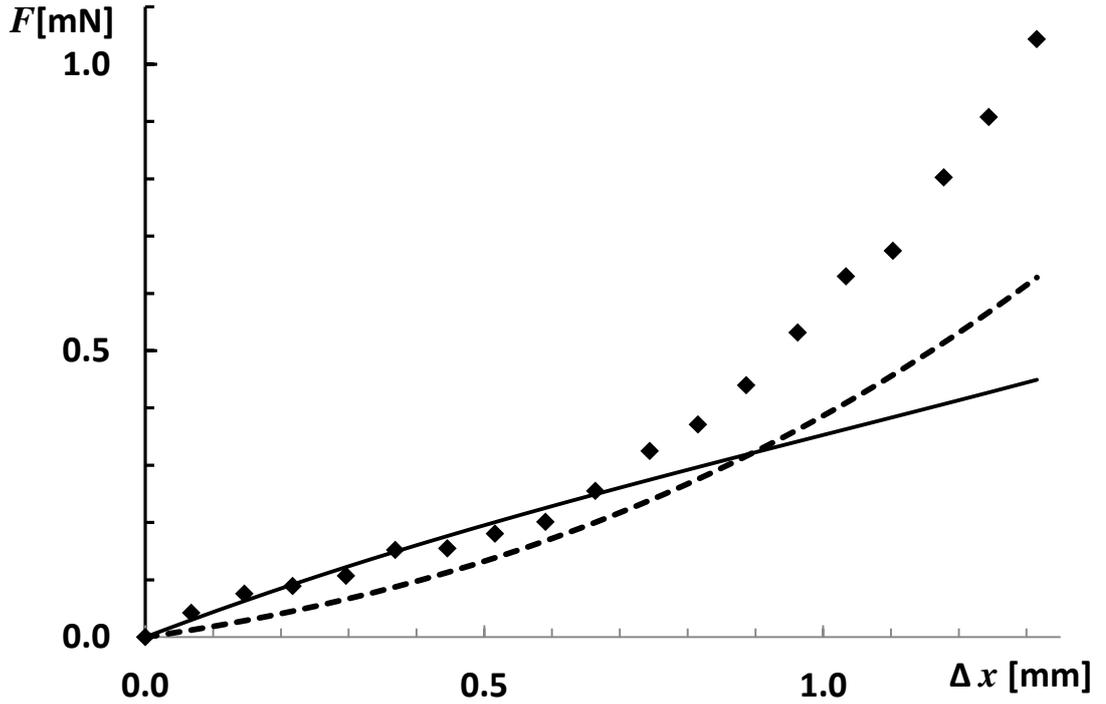

**Fig. 3.** Comparison of the elastic force calculated according to (6) (solid line) for the marble shape shown in Fig. 2A with the experimental data of Ref. [37] (diamonds). The effective surface tension of a liquid marble is put equal to that of water, $\gamma_{eff} = 72 \text{mN/m}$. The diameter deformation $\Delta x$ is twice as large as the radius deformation $h$ used in the text. The volume of the polyethylene covered marble was 15μL. The dashed line relates to the calculation according Eq. (10) for the marble shape shown in Fig. 2B.

For large deformations, a marble shape is more likely to a "pancake" with a profile given in Fig. 2B. Its constant volume and increase in the surface area are given by the following expressions:

$$V = \frac{4}{3}\pi R^3 + \pi^2 a R^2 + 2\pi R a^2,$$

$$\Delta S = 4\pi R^2 + 2\pi^2 R a + 2\pi a^2 - 4\pi R_0^2.$$

The parameter $a$ (see Fig. 3B) preserving the volume can be obtained to the 4-th power of the strain $\varepsilon$ as:

$$a = R_0(a_1\varepsilon + a_2\varepsilon^2 + a_3\varepsilon^3 + a_4\varepsilon^4),$$

where $a_1 = 4/\pi$, $a_2 = 4(1 - 8/\pi^2)/\pi$, $a_3 = 0.27460$, $a_4 = 0.21681$ (the expressions for $a_3$ and $a_4$ looks similar to $a_1$, $a_2$ but are somewhat cumbersome). This leads to the increase in the surface energy due to deformation $\Delta E = \gamma_{eff}\Delta S$ and to the elastic force $F = d\Delta E/dh$, expressed by:

$$F = \gamma_{eff}R_0(4.760\varepsilon + 13.55\varepsilon^2 + 14.47\varepsilon^3), \qquad (10)$$

plotted in Fig. 3. It is seen that Fig. 2B and Eq. (10) are more exact at lager deformations, while Fig. 2A and Eq. (6) are more appropriate at small deformations when the marble profile is closer to a sphere. The proposed models explain semi-quantitatively the origin of elastic properties of liquid marbles.

## 4. Conclusions

In conclusion, a simple model taking into account the growth in the marble surface and the conservation of volume explains semi-quantitatively the elastic properties of marbles in a satisfactory agreement with the reported experimental data. At small deformations, the dependence of the elastic force on the deformation undergoes an inflection point, as well as the stress-strain curve. This ascertains a natural limit for application of the Hooke law in the case of marbles. The evaluation of the effective Young modulus and its dependence on the marble volume are obtained. The explanation of elastic properties of marbles simply by the increase in their surface areas brings to a question the use of more sophisticated theories for this purpose [37, 39-40].

**Acknowledgments**

The authors are thankful to Professor R. Sedev for his kind sending the experimental results from Ref. 37 to us. The authors are thankful to Mrs. Yelena Bormashenko for her kind help in preparing this manuscript.